**Discrepancies and the Error Evaluation Metrics for Machine Learning Interatomic Potentials**

Yunsheng Liu [1], Xingfeng He [1], and Yifei Mo [1,*]

[1] Department of Materials Science and Engineering, University of Maryland, College Park, MD, USA

* Email: yfmo@umd.edu

**Abstract.** Machine learning interatomic potentials (MLIPs) are a promising technique for atomic modeling. While small errors are widely reported for MLIPs, an open concern is whether MLIPs can accurately reproduce atomistic dynamics and related physical properties in molecular dynamics (MD) simulations. In this study, we examine the state-of-the-art MLIPs and uncover several discrepancies related to atom dynamics, defects, and rare events (REs), compared to ab initio methods. We find that low averaged errors by current MLIP testing are insufficient, and develop quantitative metrics that better indicate the accurate prediction of atomic dynamics by MLIPs. The MLIPs optimized by the RE-based evaluation metrics are demonstrated to have improved prediction in multiple properties. The identified errors, the evaluation metrics, and the proposed process of developing such metrics are general to MLIPs, thus providing valuable guidance for future testing and improvements of accurate and reliable MLIPs for atomistic modeling.

## 1. Introduction.

Atomistic modeling, which simulates physical phenomena based on the interactions of atoms, is a crucial research technique in a wide range of disciplines including physics, chemistry, biology, and materials science. Density functional theory (DFT) calculation have been the standard technique for evaluating atom interactions among a diverse range of configurations and chemistries, but their applications are limited to small system sizes of a few hundred atoms (up to a few nm) due to high computation costs.[1–3] By contrast, classical interatomic potentials, also known as force fields, have significantly lower computation costs and thus can be employed for atomistic simulations with much larger length-scale (nm – $\mu$m) and longer time-scale (ns – $\mu$s), but they lack the transferability to different atomistic configurations that are not considered in the potential fitting.[1,4,5]

As an emerging technique to bridge the gaps among different computational techniques [1–3,6–10], machine learning interatomic potentials (MLIPs) utilize machine learning (ML) models to predict energies and forces of atomistic structures, which are mapped into the atomistic descriptors as input. Current state-of-the-art MLIPs include Gaussian Approximation Potential (GAP) based on Smooth Overlap of Atomic Positions (SOAP) descriptors,[11,12] Neural Network Potential (NNP),[13,14] Spectral Neighbor Analysis Potential (SNAP),[15] Moment Tensor Potential (MTP),[16] and Deep Potential (DeePMD) models,[5] and many MLIP variances derived from different modifications and combinations of ML models and descriptors. MLIPs are trained using DFT-calculated energies and forces from a diverse range of atomistic configurations, typically encompassing bulk and defected structures, equilibrium and non-equilibrium structures, and solid and liquid phases. Current state-of-the-art MLIPs are claimed to achieve accuracies similar to ab initio calculations,[1,5,11,13,15–18] while maintaining low computation costs and linear size scaling akin to classical interatomic potentials.

However, the MLIPs are black-box predictors not directly based on physical principles. An open question is whether MLIPs can always accurately reproduce physical phenomena in atomistic simulations. Conventional ML error testing primarily quantifies MLIP accuracies through average errors, such as root-mean-square error (RMSE) or mean-absolute error (MAE), of energies and atomic forces across a range of configurations known as testing dataset. These atomistic configurations in the testing dataset are randomly split from the entire datasets generated in the same manner as the training dataset, and thus are similar to the training dataset but may differ from the atomistic configurations that may encounter during MD simulations. Most MLIPs are reported to achieve small, average errors of energies and atomic forces as low as 1 meV atom$^{-1}$ and 0.05 eV Å$^{-1}$,[1,5,8,11,19,20] respectively, in conventional ML testing. These low averaged errors reported have created the impression that MLIPs are as accurate as DFT calculations. However, these MLIPs with small average errors may not always accurately reproduce the physical phenomena in atomistic simulations, as shown in the following examples.[10,21,22] An MLIP of Al by Botu *et al.* was reported a low MAE force error of 0.03 eV Å$^{-1}$, but its MD simulations predicted the activation energy of Al vacancy diffusion with an error of 0.1 eV compared to the DFT value of 0.59 eV, even though vacancy structures and vacancy diffusion were included in the training dataset.[23] In Vandermause et al. [20], an Al MLIP with a low RMSE force error of 0.05 eV Å$^{-1}$ for solid phase and 0.12 eV Å$^{-1}$ for liquid phase also exhibited discrepancies with DFT in in surface adatom migration, which were considered during the on-the-fly training. Zuo et al.[1] reported a number of MLIPs (such as GAP, NNP, SNAP, MTP) with small RMSEs of atomic forces at the level of 0.15 - 0.4 eV Å$^{-1}$ and 10-20% errors in the vacancy formation energy and migration barrier for a number of materials (such as Li, Mo, Si, Ge), while vacancy structures were also included in the training. Additionally, in the aforementioned studies, there were large errors of the predicted migration energy barriers of defects, even though these defects and their migrations were considered in the training and testing datasets. Atomistic diffusion is determined by the dynamics of atoms (often in the form of point defects) and the PES beyond their equilibrium sites, but the direct testing of MLIPs on these atomistic-level details in MD simulations still shows errors and failures.[24] As reported by Fu et al.[24], the MD simulations based on MLIPs observe errors, such as radial density functions, and even the failure of the MD simulations after a certain duration. These results suggest there are errors in the MLIPs causing these errors and failures from actual MD simulations. It is crucial to examine the accuracy of MLIPs in simulating the atomic dynamics and reproducing physical properties, understand potential discrepancies, and develop appropriate testing metrics.

Typical approaches to improve the MLIPs include adjusting the fraction or weights of certain structures in the training dataset, modifying the cost/loss functions, and tuning hyperparameters.[25] The average errors in energies and forces or a few easily computable properties, such as elastic constants, energy vs. volume curves for different crystal structures, and formation energies of point defects,[1,26] are often used to optimize and select the MLIP models. However, to test and quantify those errors that can only be directly observed in actual MD simulations, such as the errors in diffusional properties, one would need to conduct numerous tests in MD simulations for an extended duration before selecting the final MLIP models.[24] This approach requires a large computational cost of running MD simulations to select MLIPs, which may be impractical for optimizing MLIPs with many combinations of training variables and hyperparameters. Therefore, appropriate testing metrics should be developed to thoroughly gauge the ability of an MLIP in reproducing atomic dynamics and physical properties in a range of typically encountered physical situations, and such quantitative metrics are crucial for the further improvement of MLIPs.

In this study, by comparing atomic dynamics from MLIP-MD simulations and ab initio MD (AIMD) simulations, we reveal that state-of-the-art MLIPs, even with carefully selected training datasets and small average errors evaluated by conventional testing, may not fully reproduce atomic dynamics or related properties (Section 2.1). The tested MLIPs show discrepancies in diffusions or rare events (Section 2.2), defect configurations (Section 2.3), and atomic vibrations (Section 2.4). We then develop the error evaluation metrics based on the atomic forces of RE atoms (Section 2.5) and demonstrate them for indicating the performance of MLIPs on atomistic dynamics in MD simulations (Section 2.6). The MLIPs trained with enhanced RE data and selected by newly developed metrics show improved predictions of atom dynamics and diffusional properties. In the end, we summarize our process of developing the evaluation metrics for the observed simulation-based discrepancies. The identified discrepancies, our evaluation metrics, and their development process are general to all MLIPs and can serve as guidance for future development and improvements of accurate, robust, and reliable MLIPs for atomistic modeling.

**2. Results**.

***2.1. Discrepancies of MLIPs on atomic dynamics between MLIP and DFT occur even when low average errors are reported***

We conduct our study on a number of MLIPs to show the observed phenomena are general among MLIPs. In order to perform a consistent comparison, we directly retrieve the MLIP (GAP, GAP$_{PRX}$, NNP, SNAP, and MTP) models of Si from previous studies[1,26], besides the DeePMD model [27] trained using the same training dataset from Ref. [1]. This training dataset of Ref. [1] includes a diverse range of selected atomistic structures of solid Si, melted liquid Si, strained Si, Si surfaces, and Si-vacancy from AIMD simulations at a wide range of temperatures (Fig. 1c, Methods and Supplementary Note 1). In order to quantify the errors on predicted energies and atomic forces of these MLIPs, we construct interstitial-RE and vacancy-RE testing sets, $\mathcal{D}_{RE-I}^{Testing}$ and $\mathcal{D}_{RE-V}^{Testing}$, respectively, each consisting of 100 snapshots of atomic configurations with a single migrating vacancy or interstitial, respectively, from AIMD simulations at 1230 K (Fig. 1a and b, Methods), with the true values of energies and atomic forces evaluated by DFT calculations with a **k**-point mesh of 4×4×4 (DFT K4). The MLIPs accurately predict the energies and atomic forces in the training dataset and the testing dataset in consistency with the original studies,[1,26] showing low root-mean-square errors (RMSEs) below 10 meV atom$^{-1}$ for energies and 0.3 eV Å$^{-1}$ for forces (with the median force magnitude of 1.67 eV Å$^{-1}$) for most MLIPs on the vacancy-RE testing set $\mathcal{D}_{RE-V}^{Testing}$ (Fig. 1e, h, Supplementary Figure 1, and Supplementary Table 1). Given similar structures with vacancies are covered in the training dataset, the good performance of MLIPs on vacancy structures should be expected.

We also test the MLIPs on the structures not included in the training data (Fig. 1c), specifically the interstitial-RE testing dataset $\mathcal{D}_{\text{RE-I}}^{\text{Testing}}$, which comprises the snapshots from the AIMD simulations of Si supercell with an interstitial. Most MLIPs (except for GAP$_{\text{PRX}}$) do not include the configurations with an interstitial in the training dataset. While the MLIPs prediction of atomistic energies show a bias with an average offset of 10 – 13 meV atom$^{-1}$ lower than DFT values, the overall RMSEs are below 15 meV atom$^{-1}$ and 0.3 eV Å$^{-1}$ (with the median force magnitude of 1.69 eV Å$^{-1}$) for these interstitial structures for most MLIPs (Fig. 1d, g, Supplementary Figure 2, and Supplementary Table 1). These small error values in energy and force on training and testing datasets are often interpreted as good performance of MLIPs.

In addition to the errors of energies and forces, we evaluate and compare the phonon dispersion relations predicted by MLIPs with DFT (Fig. 1f, i, Supplementary Figure 3, and Supplementary Figure 4). While most MLIPs have good agreement on phonon dispersions of bulk Si (Fig. 1f)[26,28], a number of MLIPs exhibit noticeable differences on phonon dispersions in the Si supercell with a vacancy (Fig. 1i and Supplementary Figure 4, Methods). For example, the phonon dispersion by GAP has imaginary frequencies (Fig. 1i), and the phonon dispersion by SNAP has additional bands with lower frequencies compared to DFT K4 in high frequency sections (Supplementary Figure 4). The training data includes the atom vibration from AIMD simulations with vacancies, which should help the training of phonon dispersion relations. Therefore, these discrepancies require further investigation.

## 2.2 Rare events are sources of discrepancies

Here the MD simulations using these MLIPs are performed to simulate atom dynamics and related physical properties in a Si supercell with a single vacancy or a single interstitial, to identify potential discrepancies between MLIP-MD and AIMD simulations. The Si diffusivities are evaluated at a range of temperatures 730 – 1600 K from the mean-squared-displacements of Si over time (Methods) (Fig. 2a and b). Given the MLIPs are trained on the DFT calculations based on a fine **k**-point mesh of 4×4×4 (K4), the MLIP results should be compared to DFT K4 benchmark as in many of our tests. For the MD simulations, the MLIP results are compared to the AIMD simulations based on coarser accuracy setting of a **k**-point mesh of 2×2×2 (K2) and a single $\Gamma$-point (K1), because AIMD simulations based on K4 is takes prohibitively long to obtain enough number of atom hops. In addition, these lower accuracy settings are commonly utilized for AIMD simulations in previous studies.[4,29,30] The errors with lower-accuracy DFT calculations also can be used as error ranges for comparison.

For vacancy diffusion, which is covered in the training data, most MLIPs predict diffusivities within a reasonable error range of the diffusivities predicted by K2 AIMD simulations. Most MLIPs perform better than DFT-K1. DeePMD gives higher diffusivities than AIMD simulations, but an agreement on the fitted activation energy (0.17 eV compared with 0.2 eV given by AIMD K2). Some MLIPs show discrepancies and deviations among each other in the diffusivities at lower temperatures, leading to discrepancies in the fitted activation energies and extrapolated room temperature diffusivities. Nonetheless, the comparison of the MLIPs and AIMD simulations should take into consideration the stochastic nature of estimating diffusivities from MD simulations and the limited number of data points, which would lead to large uncertainty of the fitted activation energy and extrapolated diffusivity.[4] As shown in the previous study[4], even for a total mean square displacement of ion diffusion over a few thousand Å$^2$, the standard deviation of obtained diffusivity can be as large as 20-30%. The error bars and range of all obtained diffusivity values are shown in Supplementary Table 4, 5 and 6.

While the Si interstitials are not covered in the training of MLIPs (except for GAP$_{PRX}$), all MLIPs show low RMSEs in the testing dataset (Fig. 1d, g) with interstitial configurations. Here the MD simulations of interstitial diffusion is performed to test whether the atomic dynamics can be correctly reproduced. Some MLIPs, such as GAP$_{PRX}$, DeePMD, and SNAP, show significant deviations in the interstitial diffusional properties predicted by MLIP-MD simulations (Fig. 2a and Supplementary Table 2). The diffusivities predicted by GAP and MTP agree reasonably with AIMD simulations over the temperature range 1000 – 1600 K, and the fitted activation energies of interstitial diffusion, $E_a^I$, show minor differences. GAP$_{PRX}$, which is the only MLIP considered interstitial in the training, also shows discrepancies in interstitial diffusivities with a low $E_a^I$ of 0.12 eV. SNAP shows a $E_a^I$ of 0.74 eV, much higher than AIMD K2 (0.30 eV). For NNP, the crystalline Si structure melted during the MD simulations at the temperature of 730 – 1600 K (Supplementary Note 4). According to the tests, many MLIPs show some discrepancies in predicting diffusivity, activation energy, or both, and thus the ability to fully reproduce the atom dynamics of Si interstitial diffusion is limited. The discrepancies in diffusional properties indicate that having low average errors in energies and forces are insufficient error evaluation metrics to judge whether the atomic dynamics of diffusions in MD simulations are accurate.

### 2.3 Configurations with similar energies are discrepancy sources

In order to reveal the discrepancies of MLIPs in predicting interstitial diffusions, we analyze the snapshots from MLIP-MD simulations of interstitials in comparison with AIMD simulations. DFT studies[31–33] reported three types of Si interstitials, such as the ground-state split-<110> (Fig. 3a), tetrahedral (Fig. 3b), and hexagonal (Fig. 3c) interstitials, with the formation energies $E_f$ of 3.56, 3.72, and 3.59 eV, respectively, from DFT (K4) calculations (Fig. 3). Consistent with the trend of the formation energies, the split-<110> interstitial has higher occurrence frequency in the AIMD simulations than tetrahedral or hexagonal (Methods) (Fig. 3d and e).[31–33] By contrast, MLIP-MD simulations show higher occurrence frequencies of either tetrahedral interstitial (by GAP, SNAP, and MTP) or hexagonal interstitial (by DeePMD) (Fig. 3e and Supplementary Figure 7). Consistent with the occurrence frequencies in MLIP-MD simulations, most MLIPs (except for GAP$_{PRX}$) give lower $E_f$ for the tetrahedral interstitial than the ground-state split-<110> interstitial (Fig. 3d). The trend of increasing occurrence frequencies for different interstitials can be largely explained by the decreasing formation energies, while the entropy of these defects may also play an effect (Supplementary Note 5 and Supplementary Note 10). Therefore, the errors in interstitial diffusion during MLIP-MD simulations are caused by the discrepancies of MLIPs in the predictions of different interstitial configurations. These discrepancies should be expected given interstitial configurations are not considered in the training (except for GAP$_{PRX}$). However, it is worth noting that small errors are reported for the testing of energies and forces for interstitial configurations (in the interstitial-RE test set $\mathcal{D}_{RE-I}^{Testing}$). Thus, having small average errors in energies and forces in conventional error testing on the defect configurations may be insufficient in determining whether the MLIPs would correctly reproduce these defect configurations in MD simulations.

### 2.4 Vibration near defects is a discrepancy source

The good performance of MLIPs on Si vacancies is expected given the dynamical snapshots of Si vacancies in a wide range of conditions are well covered in the training dataset. However, the phonon dispersion of the vacancy structure predicted by the MLIPs (Fig. 1i) suggests potential discrepancies of potential energy surface (PES) and atomic vibrations. Additionally, for the vacancy diffusion (Fig. 2a, Supplementary Table 3), the discrepancies of diffusional properties are observed in a few cases (Supplementary Note 4). The diffusional properties, such as activation energies and the pre-exponential factors of the Arrhenius relation of diffusivity, are critically dependent on the PES of a migrating atom next to a defect such as a vacancy.

We further analyze the vibrations of Si atoms neighboring the vacancy, which is directly related to vacancy diffusion, in order to gain insights into the observed discrepancies of the PES. We visualize the vibrations of Si atoms near the vacancy by plotting the distribution of the distance $r_s$ and $r_v$ of the atoms from the nearest static Si site to its nearest static vacancy site, respectively (Fig. 4a) (Methods). For the atoms that are not the nearest neighbor (NN) atoms of the vacancy (Fig. 4d), the atom vibrations are similar to those in crystalline bulk. The agreement in the vibration of non-NN atoms is consistent with the good agreement in the phonon dispersion relations (Fig. 1f) and elastic constants of bulk crystalline Si.[1,26,28]

The vibrations of the vacancy nearest neighbor (vacancy-NN) atoms in MLIP-MD simulations show major discrepancies with AIMD simulations (Fig. 4e – h, and Supplementary Figure 8). In the MLIP-MD simulations, the vacancy-NN atoms vibrate much further away from their static sites, as indicated by the distributions of $r_s$ (Fig. 4c, Supplementary Figure 8). The discrepancies in vacancy-NN vibrations reveal that MLIPs do not accurately reproduce atom dynamics around the defect.

The observed discrepancies in the vibrations of vacancy-NN atoms can be explained by the errors of the PES of a vacancy-NN Si atom moving along the direction toward the vacancy (Fig. 4b and Supplementary Figure 10). While some pathways of the migration for a single atom in a static configuration with fully relaxed static sites (Supplementary Figure 10) may be poorly predicted by the MLIPs, some different pathways can be accurately predicted in very similar configurations with slightly adjusted atom positions (near-equilibrium snapshots from AIMD simulations) (Supplementary Note 6, Supplementary Figure 11). These observed discrepancies indicate that the PES predicted by the MLIPs may not be always accurate and reliable under similar atomic configurations with minor variances in positions.

These errors in the atom vibrations and PESs nearby a defect are surprising, given that a range of dynamical configurations of vacancies from AIMD simulations is considered in the training data and that small errors in energies and forces are reported for these configurations. These results show the challenges of MLIPs in accurately reproducing the PES of atoms related to defects, even if these defects and related dynamics are included in the training. In addition, the conventional error testing of MLIPs based on average errors of energies and forces are insufficient in indicating these errors in atom vibration and PES near defects. Our results reveal previously neglected errors on the PES of MLIPs in causing inaccurate atom dynamics.

In summary, we have identified discrepancies based on observations of atomic dynamics in MD simulations, such as diffusions, configurations of defects, and atomic vibrations. In order to train MLIPs that can more accurately reproduce these dynamical phenomena, we need to quantify these discrepancies by developing corresponding error evaluation metrics, which can be further used to train and select the MLIPs with the highest metric scores.

### *2.5 Quantifying the force errors on RE migrating atoms*

Using the discrepancies on atom diffusions as an example, we here develop corresponding error evaluation metrics, and improve the performances of MLIPs on diffusional properties. The process is as follows. We first develop a number of metrics for quantifying the aforementioned sources of discrepancies (Section 2.5). The evaluation metrics are then statistically verified to effectively indicate diffusional properties derived from atom dynamics in MD simulations (Section 2.6). This process can be generalized to improve the training and testing of MLIPs, as summarized in Section 2.7.

Given the aforementioned discrepancies in diffusional properties, we identify the sources of errors on migrating atoms, which are the atoms of interstitial or vacancy in the middle of the hopping from the current equilibrium site to the neighboring site (Methods). These atom hops are known as rare events (REs) in MD simulations. To quantify the errors of MLIP predictions for RE atoms, we compare the predicted atomic forces with DFT results on these RE atoms. We evaluate the error of atomic force in the magnitude $\delta_F$ and the direction $\delta_\theta$ of as

$$\delta_F = |\vec{F}_{\text{predicted}}| - |\vec{F}_{\text{DFT}}| \qquad \text{Eq. (1)}$$

$$\delta_\theta = \arccos(\frac{\vec{F}_{\text{predicted}} \cdot \vec{F}_{\text{DFT}}}{|\vec{F}_{\text{predicted}}| \cdot |\vec{F}_{\text{DFT}}|}) \qquad \text{Eq. (2)}$$

where $\vec{F}_{\text{DFT}}$ is the benchmark true force values calculated by DFT K4, and $\vec{F}_{\text{predicted}}$ is the force predicted by the MLIPs (or DFT K1, or K2). For the RE atoms in the $\mathcal{D}_{\text{RE-I}}^{\text{Testing}}$ and $\mathcal{D}_{\text{RE-V}}^{\text{Testing}}$ datasets, large errors were found in the force magnitude $\delta_F$ (Fig. 5a) and force direction $\delta_\theta$ (Fig. 5f) of the atomic forces predicted by MLIPs. For the RE atoms and the nearby atoms within a distance $r < 3$ Å, 50% - 80% and > 40% exhibit large errors of $\delta_F > 0.5$ eV Å$^{-1}$ or $\delta_\theta > 15°$, respectively, whereas around 20% of the other atoms that are > 3 Å away from the migrating atoms show similar levels of errors (Supplementary Note 7, Supplementary Table 7). While other factors in addition to the force errors on RE atoms may contribute to the discrepancies, the analyses nonetheless confirm that there are large errors on the atoms near defects and RE atoms are a major source responsible for discrepancies in diffusions.

The cumulative distribution functions (CDFs) of $\delta_F$ and $\delta_\theta$ on RE atoms show the distributions of force errors of MLIPs (Fig. 5d, e, i, j, and Supplementary Figure 12). Most MLIPs (except for GAP$_{\text{PRX}}$) exhibit errors in force prediction on RE atoms (8-25% for $\delta_F > 0.5$ eV Å$^{-1}$ and 40-70% for $\delta_\theta > 15°$), whereas lower accuracy DFT K2 (< 0.5% for $\delta_F > 0.5$ eV Å$^{-1}$ and 1% for $\delta_\theta > 15°$) and DFT K1 calculations (5% for $\delta_F > 0.5$ eV Å$^{-1}$ and 30% for $\delta_\theta > 15°$) have much smaller errors. These results clearly show MLIP predictions have large errors on RE atoms. The force magnitude error $\delta_F$ of larger than 0.5 eV Å$^{-1}$ is significant, as the median force magnitude of all atoms is around 1.7 eV Å$^{-1}$ (Fig. 5b). Among these atoms with $\delta_F$ > 0.5 eV Å$^{-1}$, 10 – 35% interstitials and 3 – 15% vacancies also exhibit significant force direction errors of $\delta_\theta$ > 30⁰ (Fig. 5g), which would lead to major errors in predicting atom dynamics in MD simulations. Therefore, these large errors in MLIP-predicted forces on RE atoms (Fig. 5d, e, i, and j) and their nearby atoms (Supplementary Table 7 and Supplementary Figure 12) are major sources of the observed discrepancies in the atom vibration and diffusion between MLIPs and DFT.

By identifying what atoms tend to give large errors, it can be understood why conventional error testing of MLIPs shows very small errors. In conventional error testing of MLIPs, most of the atomistic configurations evaluated are for atoms near equilibrium positions, which are accurately predicted by MLIPs with very small errors. The RE atoms only consist of a very small fraction in most typical testing datasets, e.g., < 1% in the dataset of Zuo et al.[1] and Bartok et al.[26] (2 – 3% in our RE-testing datasets). Therefore, the large errors on RE atoms are averaged out by the majority of near-equilibrium atom configurations. The total RMSEs of the forces on all atoms in conventional error testing mostly reflect the accurate MLIP prediction of near-equilibrium atoms, which often dominate the typical testing dataset (Supplementary Note 7). However, it should be noted that a few percent of RE-related atoms are critical for the correct prediction of atom dynamics and physical phenomena in MD simulations.

In order to overcome the limitations of conventional error testing, we propose to quantify the CDFs of force errors $\delta_F$ or $\delta_\theta$ on RE atoms, using the normalized area of curve (NAC) under the CDF curves (see details in Methods) as new metrics. The NAC of CDF equals to 1 for an MLIP that completely agree with the benchmark true values (DFT

K4), and a lower value of NAC suggests larger errors. To combine the metrics of force magnitude $NAC(\delta_F, \mathcal{D})$ and force direction $NAC(\delta_\theta, \mathcal{D})$, we define the force performance score $P(\mathcal{D})$, as the product of $NAC(\delta_F, \mathcal{D})$ and $NAC(\delta_\theta, \mathcal{D})$ for errors of atomic forces in a given dataset $\mathcal{D}$ (Methods),

$$P(\mathcal{D}) = NAC(\delta_F, \mathcal{D}) \times NAC(\delta_\theta, \mathcal{D}). \qquad \text{Eq. (3)}$$

These quantitative metrics can be effectively used in training, validation, and testing to improve MLIPs as demonstrated in the next section.

### *2.6 Force performance scores as effective metrics*

To evaluate the effectiveness of $P(\mathcal{D})$, we here obtain MLIPs with high $P(\mathcal{D})$ scores and compare their atomic dynamics in diffusions with previous MLIPs. A RE-enhanced training dataset is generated to train the MLIPs that can achieve higher $P(\mathcal{D})$ scores. For the RE-enhanced training dataset, we replace a fraction (54%) of the structures in the original dataset in Ref.[1] by those containing identified migrating interstitials. In this way, the size of the training dataset is kept the same to eliminate the impact of training data size in the comparison with original MLIPs (Methods). In the RE-enhanced validation, the force errors on RE atoms are evaluated for trained MLIPs using the enhanced validation set (EVS) (Methods), which includes those structures with RE atoms (migrating interstitials and vacancies) identified in addition to a fraction of the original dataset in Ref.[1]. We use the evaluation metrics on errors of energies, overall forces, forces of RE atoms (migrating interstitials or vacancies) to fine-tune the hyperparameters of MLIPs and select the MLIPs with good performances on all evaluation metrics in the validation process (Methods). Following this process, we train and obtain 135 MLIPs to study the statistical effectiveness of our force performance score and six representative RE-enhanced MLIPs (denoted by the subscript RE-I), GAP$_{RE-I}$, NNP$_{RE-I}$, DeePMD$_{RE-I}$, SNAP$_{RE-I}$, and MTP$_{RE-I}$ and compare their atomic dynamics in MD simulations with previous MLIPs (Methods).

To evaluate the effectiveness of our force performance scores, here we calculate and compare the force performance scores $P(\mathcal{D})$ and the diffusional properties from MD simulations, for a total of 81 MLIPs with a range of models and different hyperparameters, such as 35 original MLIPs and 46 interstitial-enhanced MLIPs selected by the metrics of force performance scores $P(\mathcal{D})$ in the validation step (Methods). Eight criteria measuring the energy errors and force errors of both force magnitudes and directions (Methods) on RE atoms are evaluated. The diffusional properties, such as activation energies, diffusivities, and diffusion pre-factors, from MD simulations of RE-enhanced MLIPs with high force performance scores, show improved agreement with AIMD simulations (Fig. 6b). Among these 81 MLIPs (Fig. 6a), the MLIPs with higher force accuracies $P(\mathcal{D}_{RE-I}^{Testing})$ on interstitial RE atoms than DFT K1 predict the activation energy $E_a^I$ of interstitial diffusion in good agreement with the AIMD K2 value (Fig. 6a and Supplementary Figure 14) (Supplementary Note 9). The other MLIPs with lower $P(\mathcal{D}_{RE-I}^{Testing})$ lower scores than DFT K1 (below the red dashed line in Fig. 6a) show much large variations and errors in their predictions of diffusion. Among the RE-enhanced MLIPs with high performance scores, GAP$_{RE-I}$, SNAP$_{RE-I}$, NNP$_{RE-I}$, and MTP$_{RE-I}$, give $E_a^I$ of 0.2, 0.33, 0.29, and 0.26 eV, respectively, in better agreement with 0.25 and 0.30 eV from K1 and K2 AIMD simulations, compared to 0.42, 0.74, and 0.42 eV by the original MLIPs, (Fig. 6b, Supplementary Table 2). The predicted diffusion pre-factors $D_0^I$ of enhanced MLIPs improve from >5.7×10$^{-5}$ cm$^2$s$^{-1}$ (except for GAP$_{PRX}$) in original MLIPs to 8.5×10$^{-6}$ - 4.1×10$^{-5}$ cm$^2$s$^{-1}$, in better agreement with $7.7 \times 10^{-6}$ - $2.3 \times 10^{-5}$ cm$^2$s$^{-1}$ from AIMD simulations (Fig. 6b). In addition, the interstitial configurations observed in MLIP-MD simulations also agree better with AIMD simulations. The occurrence frequencies of split-<110> interstitials increase to around 35% in RE-enhanced GAP$_{RE-I}$, NNP$_{RE-I}$, SNAP$_{RE-I}$, and MTP$_{RE-I}$ from less than 10% by the original MLIPs (Fig. 3e). In addition to these improvements, the occurrence frequencies of ground-state split-<110> interstitial are still lower than those of ground-state tetrahedral, even the configurations of all three types of interstitials are included in the RE-enhanced training data. Given the total energy difference of 0.16 eV between the two interstitial configurations is merely 3 meV atom$^{-1}$ in the 65-atom supercell used, these results indicate the difficulties of MLIPs in accurately reproducing the atomic configurations with similar energies. In summary, the force performance scores on RE atoms are effective error evaluation metrics to indicate related diffusional properties predicted by the MD simulations of the MLIPs. Using improved error evaluation metrics for optimization and validation is demonstrated as an effective step in training MLIPs with improved performance.

## *2.7 Process of developing error evaluation metrics*

Developing error evaluation metrics that are indicative of the predictions of atomic dynamics is essential for the development of MLIPs. In the conventional training process, optimizing MLIPs on properties other than the errors of predicted energies and forces can be either done by 1) adding additional terms and weights into the loss functions as training targets (Fig. 7a-i),[1,25] or 2) adding additional metrics or material properties, such as elastic tensors in Ref. [1], when deciding the optimal hyperparameters for MLIPs (Fig. 7a-ii). However, this conventional approach would be computationally expensive for evaluating atomic dynamics (Fig. 6a-ii), because it requires running multiple MD simulations for each trained model for the fine-tuning of hyperparameters.[24] Thus, effective error evaluation metrics that do not require extensive MD simulations and are indicative of atomic dynamics are essential, so one can train MLIPs that can accurately predict physical phenomena in MD simulations with lower computational costs in the training and validation process.

Here we summarize the process of developing error evaluation metrics as follows, which can be generalized for future development to improve simulation-based performance of MLIPs (Fig. 2b): (1) Identify the sources of discrepancies in the MD simulations, (2) propose related but easy-to-calculate metrics that are related to the physical properties based on the simulation (e.g. forces on RE atoms to diffusional properties in our case), (3) quantify the proposed error evaluation metrics and dynamical properties based on simulations for a range of MLIPs, and (4) verify the effectiveness of proposed metrics statistically among these MLIPs with different models, hyperparameters, or training data. If the metrics prove to have a statistically significant effect on improving the atomic dynamics for MLIPs, we then can add the proposed and verified evaluation metrics into the training process (Fig 7a-ii).

There are important considerations in generalizing this process of developing error evaluation metrics for simulation-derived properties. It's important to use a large amount of data, on metrics and dynamical properties, predicted by diverse MLIPs, and conduct statistical verification instead of using a single or a few MLIPs because the developed metrics should be general enough to be applied to most MLIPs, which may be in the future trained by other descriptors, models, or training data. In some exception cases, MLIPs (e.g., NNPs (orange crosses), SNAPs (purple crosses), and the other models (circles) in Fig. 6a) may give poor scores in evaluation metrics but may have small errors on dynamical properties, thus statistically verifying the evaluation metrics is important. There are many factors affecting the outcome of the training of ML models, and many are random in nature and yet poorly understood.

## *2.8 Trade-offs in MLIP performances: Pareto fronts of MLIPs*

Additionally, we compare the performance and force accuracies of MLIPs based on a variety of ML algorithms and atomic descriptors and observe the trade-off the accuracies on different properties (Fig. 6c). We compare MLIPs including the 46 interstitial-enhanced MLIPs which used training data with both interstitials and vacancies, and 54 vacancy-enhanced MLIPs trained using vacancy REs data (Supplementary Note 3), 11 original MLIPs, and 24 MLIPs retrained from the original dataset.[1] We compare the force accuracies $P(\mathcal{D}_{RE-I}^{Testing})$ and $P(\mathcal{D}_{RE-V}^{Testing})$ on interstitial- and vacancy-RE atoms, respectively. For all MLIPs, there is a clear trend of decreasing $P(\mathcal{D}_{RE-I}^{Testing})$ with increasing $P(\mathcal{D}_{RE-V}^{Testing})$ as shown in Fig. 6c, indicating trade-offs on force accuracies of different defects (interstitials versus vacancy). The Pareto front lines for each type of MLIPs (GAP, NNP, SNAP, MTP, DeePMD, and DeepPot-SE) can be shown (Fig. 6c) for the force accuracies $P(\mathcal{D}_{RE-V}^{Testing})$ and $P(\mathcal{D}_{RE-I}^{Testing})$ on different defects, such as interstitial-RE and vacancy-RE atoms.

These Pareto fronts indicate the difficulties of MLIPs in achieving accurate predictions for all properties and serve as a good way to compare the performance of different MLIPs. Some MLIPs have lower Pareto fronts, indicating low force accuracies in interstitials at given force accuracies in vacancies. Notably, a few MLIPs, such as GAP and MTP, are able to achieve higher force accuracies than DFT K1 in both interstitial and vacancy. These MLIP models and descriptors may be more effective in accurate prediction for different defect configurations at the given training dataset, though the trade-offs between different predicted properties generally exist for all MLIPs.

There are more examples showing the trade-offs of MLIPs on other properties. MTP gives accurate predictions for atomic forces and activation energies but shows discrepancies in the predicted atomic vibrations near vacancy (Fig. 3f, Fig. 6a, and c). GAP has good predictions on diffusivities in MD simulations but shows discrepancies on phonon spectra (Fig. 1f, g, and h). DeePMD reproduces activation energies of diffusion but shows errors for diffusivities, atomic forces, and interstitial configurations (Supplementary Table 2, Supplementary Table 3, Fig. 1, Fig. 2, Fig. 4, and Supplementary Figure 12). Therefore, the performance of MLIPs should not be judged by a single property or a few properties. Even if an MLIP may give good performance for a number of properties, good performance for other properties should not be assumed.

Furthermore, overcoming the trade-off of MLIPs' performance on different properties is required to further improve MLIPs for a wide variety of physical simulations. While a careful selection and balance for different types of defect data are known to be essential, it is also important to have a systematic process and quantitative metrics to train MLIPs with balanced accuracies in a range of structures and properties.

**3. Discussion**.

Our study presents a systematic testing on the current state-of-the-art MLIPs, regarding the discrepancies in the prediction of atom dynamics and physical properties, in comparison with DFT calculations. Even with the carefully selected training dataset and very low average errors in energies and atomic forces from conventional ML testing, MLIPs may exhibit significant discrepancies in their calculated properties and atom dynamics in MD simulations. Examples of these discrepancies include different defect types, atom vibration near defects, phonon dispersion, and diffusional properties. These discrepancies of the MLIPs in actual MD simulations should not be neglected. While the discrepancies for cases that are not included in the training data may be expected, significant discrepancies for presumed cases that are fully covered in the training dataset (e.g., atomic vibration around vacancy) are also significant in some cases and should be carefully tested. These sources of potential discrepancies are summarized as follows and should be carefully considered for improved training and rigorous testing in the future development of MLIPs.

*The properties related to rare events*. Atomistic diffusion is a typical example of physical property derived from MD simulations. However, the diffusion is largely determined by the REs, which are often less sampled in the training and testing data of the MLIPs.[34,35] The MLIP predictions on the atomic forces (both in force magnitude and force direction) of RE atoms deviate significantly from DFT values, causing errors in the predicted atomic dynamics and diffusional properties from the MD simulations. On the other hand, these errors associated with RE atoms serve as effective metrics for quantifying errors of MLIPs. Similar discrepancies are expected for diffusions in other materials systems or other types of rare events, such as reactions and state transitions. In addition, while the RE atoms analyzed in this study are identified using a hand-designed algorithm based on local atom distances, unsupervised machine learning, e.g., the *k*-means clustering method, as we demonstrated in the SI, can also be employed to identify these RE atoms.

*The defects with similar energies.* As shown in the examples of Si interstitials, the formation energies and the occurrence in MD simulations of different defect configurations (e.g., split-<110> vs tetrahedral interstitials) may be incorrectly predicted by MLIPs. Even if these various defects are considered in the training, as shown in our results generated by RE-enhanced MLIPs (Fig. 6b), the MLIPs' predictions still do not completely agree with the DFT benchmark. These discrepancies are not revealed by low average errors on energies and forces in conventional ML testing for MLIPs, even if the testing dataset includes these defect configurations.

*Atomic vibration.* The dynamics of atoms near defects are poorly reproduced by the MLIPs, even though these atom dynamics are considered in the training data. These errors are caused by the poor prediction of PES around defects. Since the PES is a high dimensional function of multi-atom configurations, the capability of MLIPs covering these configuration spaces may still have limitations, which should be further studied. It's worthwhile to note that the atomic vibration and the aforementioned formation energy of defect are persistent challenges for MLIPs in our study and are not fully resolved.

In general, these errors of MLIPs in defect energies, atom vibrations, and REs are related to defects and their nearby atoms. These aforementioned errors given by MLIPs, though yield large errors in the prediction of atom dynamics and properties in MLIP-MD simulations, are not reflected in the conventional testing of MLIPs. In conventional testing, a majority of the testing data are near-equilibrium atomic configurations, which are well described by the MLIPs, leading to low average error values (Supplementary Note 1). Those aforementioned error-prone atomic configurations, such as RE atoms and those around defects, account for a very small fraction of atomic configurations in the testing dataset and their errors are averaged out in the RMSE/MAE error values.

This understanding provides important guidance for the training, testing, and development of MLIPs. Careful considerations are required for complex defects, their surrounding atoms, and non-equilibrium structures in the training of MLIPs.[1,26] In addition to the evaluation of average error values, the accuracies and robustness of MLIPs should be carefully considered and tested on a number of key scenarios with a high likelihood of errors, including defects, atom vibrations near defects, and forces (both direction and magnitude) on RE atoms. As demonstrated in our study, the error evaluation metrics quantifying these errors can effectively improve the training, validation, and testing of MLIPs. For example, force accuracies on RE atoms are demonstrated examples of such error metrics to be used in the validation (Fig. 6a) and testing of MLIPs (Fig. 3, 6c). Furthermore, our demonstrated process of developing error evaluation metrics, which were verified over a large number of MLIPs with different models and training datasets and were applied to selecting MLIP models, can be generally applied and extended to alleviate or overcome these MLIP errors.

It is very important to develop these error evaluation metrics that are indicative of simulation-based errors. These error evaluation metrics serve as an applicable and effective method in the verification, testing, and selection of better MLIPs, because it is computationally expensive to run a large number of MD simulations for all MLIPs over an extended time duration to obtain related physical properties (such as diffusion). For these error evaluation metrics to be effective and generally applicable, it is important to statistically verify these metrics on a large number of MLIPs with different models and even different training datasets and to compare many metrics and their values with the targeted dynamical properties. Since evaluation metrics are effective in a statistical sense, applying them to different MLIP models, descriptors, and hyperparameters may show various levels of improvement. Nonetheless, developing and applying error evaluation metrics are critical to developing MLIPs with improved performance on atomic dynamics and other properties derived from MD simulations. The process we demonstrate to develop such evaluation metrics can be generalized and extended for future studies.

Applying our RE-based evaluation metrics provides significant improvements in MLIPs, and the aforementioned errors are significantly reduced but not fully eliminated, including atom vibration near defects and defect configuration occurrences since they're not directly related to RE atoms. Beyond the aforementioned issues of averaged errors and limited PES sampling, our results reveal a number of other potential challenges of MLIPs for improving performances on these defect-related errors. The defect-related errors, such as on different types of interstitials or the PES around defects, can be understood as atomistic configurations with small perturbations and small energy differences (~$10^1$ meV atom$^{-1}$). Given the defect has a low concentration in the supercell model and the total energies of the entire supercell model are used for training, it can be challenging for the MLIPs to accurately reproduce the PES around all these atomistic configurations with small differences and similar energies. Small differences in the PES would lead to significant change in the probability density, atomic forces, occurrence frequency, and dynamics of atoms. While more systematic testing and more training data may be helpful, it is possible that the descriptors and ML models can be further improved to capture the PES of these varying atomistic configurations. Given the PES is a high-dimension function of the atomic descriptor, it is possible the interpolation among the given configurations from the training dataset may be inadequate for accurately covering the entire relevant portion of PES encountered during MD simulations. Overall, to accurately reproduce all aforementioned error cases with similar energies, more studies are needed to further test and improve the atomistic descriptors, ML models, or the schemes of training and testing.

In conclusion, we study a number of MLIPs and identified a number of potential discrepancy sources in their applications. Leveraging these, we develop evaluation metrics into a process that identifies sources of discrepancies in atomic dynamics, quantifies the discrepancies, statistically verifies the effectiveness of the developed metrics, and optimizes MLIPs using enhanced quantitative metrics. By proposing and demonstrating improved evaluation metrics and the general process to develop such metrics, we show the improvement of the MLIPs in the prediction of physical properties. Overall, our results highlight general guidance and potential challenges in the future development of accurate, robust, and reliable MLIPs for atomistic modeling.

**Methods**

*Obtaining and Training MLIPs.* To obtain existing MLIPs in consistent with previous studies, we retrieved MLIPs (GAP, NNP, SNAP, and MTP) directly from the corresponding *mlearn* repository of Zuo et al.[1], or trained MLIPs (DeePMD) by using the training data the same as in Zuo et al.[1] The *mlearn* package and the corresponding MLIP models, including QUIP for GAP,[11] N2P2 for NNP,[36] MLIP for MTP,[16,37] and SNAP coded in LAMMPS,[38] were used for the energy and force evaluation. The GAP$_{PRX}$ model was retrieved from Ward et al.[39] Since different training data was used for the existing DeePMD model by Zhang et al.[5], the DeePMD model was trained here using the training data from Zuo et al.[1] in order to have a fair comparison with other MLIP models. Here we refer to the early version of Deep Potential Molecular Dynamics as DeePMD[5] and the later version of Deep Potential–Smooth Edition as DeepPot-SE in Supporting Information.[27] The training of DeePMD models was performed using the DeePMD-kit package.[27,40] The hyperparameters of the DeePMD model, including the size of the neural network, cut-off radius, and the number of iteration steps, were optimized using a grid search algorithm (two to six values were considered for each hyperparameter) by the RMSEs of energies and forces in a separate set of 100 snapshots from AIMD simulations of bulk Si with single vacancy defect (vacancy validation dataset). The length of the neighbor list in DeePMD model was estimated based on cutoff radius. The final set of hyperparameters was selected to have the lowest RMSEs of energies and forces in both training dataset and the vacancy validation dataset.

*First-Principles Computation.* DFT calculations were performed to generate additional data of energies, forces, PES and atomic configurations for training and testing. All DFT calculations were performed by Vienna ab initio simulation package[41] (VASP) with the projector augmented-wave approach on these snapshots. The Perdew-Burke-Ernzerhof[42] (PBE) functionals by generalized-gradient approximation (GGA) were adopted to calculate the total energies of snapshots. Static relaxation of atomic configurations were performed using spin-polarized DFT calculations with an energy cutoff of 520 eV, an electronic relaxation convergence cut-off of $10^{-5}$ eV, and other parameters set similar to those used in Materials Project.[43,44] Our benchmark true values of energies and forces were calculated using 4×4×4 **k**-point mesh (K4), while Γ-centered 1×1×1 (K1) and 2×2×2 (K2) were also evaluated for comparison.

*Ab initio molecular dynamics simulation.* Ab initio molecular dynamics (AIMD) simulations were performed using non-spin-polarized, an electronic energy convergence cut-off of $10^{-4}$ eV, a Γ-centered 1×1×1 (K1) or 2×2×2 (K2) **k**-point mesh, and a time step of 2 fs. The PBE functionals by GGA were adopted as in the *First-Principles Computation* section. The supercell model consists of Si bulk-phase with 2×2×2 conventional unit cells (64 atoms) with a single vacancy (63 atoms) or a single interstitial (65 atoms). The initial temperature of the simulations was set to 100 K after a static relaxation of the initial structures, and the structures were heated to the final temperatures during a period of 2 ps with a constant rate by velocity scaling, and afterwards an NVT ensemble with Nosé-Hoover thermostat was adopted. To obtain diffusional properties, AIMD simulations were performed at different temperatures 730, 840, 1000, 1230, 1500, and 1600 K following the same scheme in Ref.[4,45]. Missing diffusivities at certain temperatures in Fig. 2a and b are either due to the melting of the crystal structures or inadequate numbers of hopping events (specified in Supplementary Table 4, 5, and 6). Given the stochastic nature of ion hopping in estimating diffusivities, the diffusivity calculations were converged using our developed scheme.[4,45] and the error bars are estimated correspondingly based on the total number of ion hops. The diffusivities and their error bars of AIMD simulations at all temperatures are available in Supplementary Table 4, 5, and 6.

The diffusivities and their error bars were evaluated according to the total mean-squared-displacement (TMSD) of Si atoms as in Ref.[4]. The total time durations of AIMD simulations were in the range of 100 and 1000 ps, so the values of TMSD were in the range of 1200 to 3000 $Å^2$ and were 600 $Å^2$ for AIMD K2 simulations at temperatures of 1000 K and 1230 K due to high computation cost.

*MLIP-MD simulations.* All classical MD simulations based on MLIPs were carried out using LAMMPS. The MLIP-MD simulations for vacancy and interstitial diffusion were performed on the same supercells with single interstitial or vacancy defect as in the AIMD simulations with NVT ensemble. The time step of all MD simulations was set to 1 fs. MLIP-MD simulations of all MLIPs are performed at six different temperatures 730, 840, 1000, 1230, 1500, and 1600 K. The total time duration of MD simulations was in the range of 500 to 5000 ps, similar to those used in AIMD simulations. The TMSDs of MD simulations were between 6000 - 320000 $Å^2$ for NNP, SNAP, MTP, DeePMD, and DeepPot-SE, and > 1500 $Å^2$ for GAP models. The convergence of diffusivity and the error bars are estimated correspondingly based on the total number of ion hops in the same scheme as in the AIMD simulations.[4,45] The diffusivities and their error bars from MLIP-MD simulations are summarized in Supplementary Table 4 and 5, following the same scheme described in *Ab initio molecular dynamics simulation* Section.

*Analyzing Si interstitial in MD Simulations.* The occurrence frequency of each interstitial configuration (split-<110>, tetrahedral, and hexagonal) was counted among 2000 snapshots (taken every 100 fs) from the single-interstitial supercell model from MD simulations at 1230K. To determine the interstitial type, a structural matching algorithm was performed for each snapshot following the scheme adopted in Ref. [46] using *pymatgen*[43]. The interstitial configurations used as matching templates were fully relaxed using the fixed lattices of a crystalline Si bulk and a Γ-centered 4×4×4 **k**-point mesh, with energy and the force convergence criteria at $10^{-7}$ eV and 0.01 eV Å$^{-1}$, respectively. The matching algorithm used the tolerance parameters of the lattice angle of 5°, the lattice length of 20%, and the site root-mean-square tolerance of 0.275($V/n$)$^{1/3}$, where $V/n$ was the volume $V$ normalized by the number of atoms $n$. Some interstitial configurations, e.g., migrating interstitials or concerted migrations of multiple interstitials, happened in MD simulations, cannot be classified as either of three interstitials, and thus the occurrence frequencies of three interstitials do not always sum up to 1.

*Identifying Static Sites, Vacancies, and Migrating Atoms.* The algorithms for identifying static sites, vacancies, and migrating atoms during MD simulations were as follows. The static sites of Si were set to the sites of the perfect crystalline Si bulk. A static site that has no Si atoms within 1.1 Å was identified as a vacancy. A migrating Si atom (RE atom) was a Si atom between two nearest-neighbor static sites, which was selected if the distances to their 1$^{st}$ nearest and 2$^{nd}$ nearest static sites has a difference below 0.75 Å (approximately 31% of distances between two static sites). This criterion can effectively distinguish migrating atoms from atoms vibrating around static sites.

The values of the distance to the nearest static site, $r_s$, the distances to its nearest vacancy, $r_v$, and the distances to its nearest RE atom, $r$, were quantified correspondingly from these static sites. All distributions in Fig. 3c – h was generated with 50×50 grid size by collecting $r_v$ and $r_s$ distances of vacancy NN or non-NN atoms over 2000 snapshots taken randomly (> every 100 fs) from each simulation of MLIPs or AIMD at 1230 K. The Gaussian filter with smoothing parameter $\sigma$ set to 3 was applied in plotting the distributions using *scipy* library[47].

*Testing dataset with REs.* Two testing sets, $\mathcal{D}_{\text{RE-I}}^{\text{Testing}}$ and $\mathcal{D}_{\text{RE-V}}^{\text{Testing}}$ sets, are constructed from 100 snapshots taken from AIMD simulations at 1230 K with single-vacancy and single-interstitial, respectively, with every snapshot having a RE. The energies and atomic forces of each snapshot were further converged by single-step self-consistent DFT calculations to a higher accuracy (Γ-centered 4×4×4 **k**-point mesh) with fixed atom positions and lattices, and these converged values (DFT K4) were used as true values for testing.

*Error Evaluation of Atomic Forces.* Using the $\mathcal{D}_{\text{RE-I}}^{\text{Testing}}$ and $\mathcal{D}_{\text{RE-V}}^{\text{Testing}}$ sets, the error evaluation of atomic forces was performed for all MLIPs. The errors of atomic forces by DFT calculations with lower accuracy setting (K1 and K2) commonly used in AIMD simulations were also evaluated in comparison to the MLIPs. The magnitude error $\delta_f$ and the directional error $\delta_\theta$ of forces were evaluated according to Eq. (1) and (2), respectively. From the error evaluation results on the RE atoms (i.e., migrating atoms as defined above) in the $\mathcal{D}_{\text{RE-I}}^{\text{Testing}}$ or $\mathcal{D}_{\text{RE-V}}^{\text{Testing}}$ dataset, the CDFs of $\delta_F$ and $\delta_\theta$ were generated. The normalized area under the curves $NAC(\delta, \mathcal{D})$ of the CDFs of $\delta_F$ and $\delta_\theta$ were quantified in corresponding dataset $\mathcal{D}$ for $\delta_F$ over 0 – 1 eV Å$^{-1}$ (20 bins) or for $\delta_\theta$ over 0 – 90 degrees (20 bins), respectively, for each MLIP or DFT K1/K2 as shown in Fig. 4, and were normalized by the total area of the evaluation range of 0 – 1 eV Å$^{-1}$ for *x*-axis and 0 – 100% for *y*-axis, giving to a value between 0 and 1. Higher NAC values (closer to 1) correspond to smaller errors in the MLIP prediction compared to DFT K4 benchmark.

*Generating RE-Enhanced Training Dataset.* To train MLIPs with low errors of forces on RE atoms, 120 snapshots with identified RE atoms were randomly selected to replace the structures in the original training dataset[1] from the following categories, such as liquid Si, AIMD simulations of Si bulk, and the strained Si bulk, from the original training dataset, in order to maintain a balance of structures for each category. The RE-enhanced training dataset has the same size and similar diversity of different types of atomic configurations as the original dataset in Ref,[1] so we can make a fair comparison between existing and RE-enhanced MLIPs.

*Optimizing Enhanced MLIPs.* The validation dataset to optimize the MLIPs trained by RE-enhanced training data was constructed as follows. The enhanced validation set (EVS) contains 50 total structures, consisting of 20 structures randomly selected from the 120 replaced structures of the original training dataset, 11 snapshots with vacancy RE from AIMD simulations, and 19 snapshots with interstitial RE from AIMD simulations. The optimization of the hyperparameters of MLIPs considered 4 to 10 values for each parameter including the band limit of spherical harmonic basis functions and the number of radial basis functions for GAP, the size of neural network for NNP, the number of radial basis functions number for MTP. For each MLIP model, grid search algorithms were performed among 300 – 2000 sets of hyperparameters to identify the selected MLIPs as explained below.

For the selection of optimal MLIP models, which were used to study the force performances of MLIPs in Fig. 6c, we evaluated the following eight metrics, such as the RMSEs of energies and forces of 31 structures (excluding the 19 interstitial structures), the RMSEs of energies and forces of RE atoms in 11 vacancy structures, the RMSEs of the energies and forces of RE atoms in 19 interstitial structures, the RMSEs of the energies and forces of RE atoms in 30 vacancy and interstitial structures in the EVS. We used the force performance score $P(\mathcal{D})$ for the dataset $\mathcal{D}$. The evaluations were performed for different data such as all atoms in EVS, the RE atoms of the interstitial structures in EVS, the RE atoms of the vacancy structures in EVS, or all RE atoms in EVS.

Using these evaluation scores, we first selected those MLIPs with the lowest RMSE or the highest NAC in any one of the eight criteria, giving a total of 46 MLIPs as interstitial-enhanced MLIPs. The same optimization process using eight criterion was also applied on selecting additional MLIPs trained by the original dataset in Ref.[1] obtaining 24 additional MLIPs. All these 46 interstitial-enhanced MLIPs and 24 additional MLIPs were used in Fig. 6a for testing of their force performances on interstitial RE atoms. These RE-enhanced MLIPs were further down selected to have the optimal one as the representative MLIPs, using the joint force error score of $P(\mathcal{D}_{RE-I}^{EVS})^2 + P(\mathcal{D}_{RE-V}^{EVS})^2$. All NACs of $\delta_F$ and $\delta_\theta$ were computed using the RE atoms in the EVS. Six optimal MLIPs, one for each model, were selected. Both MLIP-MD simulations with single vacancy and MD simulations with single interstitial were further performed at six different temperatures to analyze their diffusional properties.

*Testing Interstitial-Enhanced MLIPs.* Our testing of the MLIPs from RE-enhanced training dataset (all 46 interstitial-enhanced MLIPs) included the evaluation of joint force error matrices calculated from $\mathcal{D}_{RE-I}^{Testing}$ and $\mathcal{D}_{RE-V}^{Testing}$ sets, $P(\mathcal{D}_{RE-I}^{Testing})$ and $P(\mathcal{D}_{RE-V}^{Testing})$ respectively, and the evaluation of their diffusivities, pre-exponential factors, interstitial occurrence frequencies and activation energies of vacancy and interstitial in MD simulations. The joint force error metrics and their calculated diffusional properties are compared as shown in Fig. 6a. The joint force error metrics, $P(\mathcal{D}_{RE-I}^{Testing})$ and $P(\mathcal{D}_{RE-V}^{Testing})$, of the existing MLIPs and the additional MLIPs were also calculated using the $\mathcal{D}_{RE-I}^{Testing}$ and $\mathcal{D}_{RE-V}^{Testing}$ sets.

**Data availability statement**

The structural (POSCAR files), energies, and forces data to support the finding of this study, including original training dataset from Ref.[1], the enhanced validation set $\mathcal{D}^{EVS}$, the interstitial-enhanced training set, the interstitial-RE testing set $\mathcal{D}_{RE-I}^{Testing}$, the vacancy-enhanced training set, the vacancy-RE testing set $\mathcal{D}_{RE-V}^{Testing}$ are available from: https://github.com/mogroupumd/Silicon_MLIP_datasets

**Code availability statement**

The computation codes and programs to support the finding of this study is available from the corresponding author on reasonable request. All DFT calculations are performed using VASP version 5.4.4. Python packages of *pymatgen*, *mlearn*, *scipy*, *quippy*, *scikit-learn*, and *DeePMD-kit* (Python interface) are used to analyze data, train corresponding MLIPs, or perform MLIP-MD simulations. MLIP-MD simulations also utilizes scripts using LAMMPS and DeePMD-kit.

**Acknowledgement**

The authors acknowledge the funding support from National Science Foundation Award# 1940166 and 2004837 the computational facilities from the University of Maryland supercomputing resources, and the Maryland Advanced Research Computing Center (MARCC).

**Author contributions**

Y.M. supervised the project. All authors designed the computation and analyses, and Y.L. performed them. Y.L. and Y. M. wrote the manuscript.

**Competing interests**

The authors declare no competing interests.

**References**

1. Zuo, Y. *et al.* Performance and Cost Assessment of Machine Learning Interatomic Potentials. *J. Phys. Chem. A* **124**, 731–745 (2020).

2. Batra, R. & Sankaranarayanan, S. Machine learning for multi-fidelity scale bridging and dynamical simulations of materials. *J. Phys. Mater.* **3**, 031002 (2020).

3. Chan, H. *et al.* Machine Learning Classical Interatomic Potentials for Molecular Dynamics from First-Principles Training Data. *J. Phys. Chem. C* **123**, 6941–6957 (2019).

4. He, X., Zhu, Y., Epstein, A. & Mo, Y. Statistical variances of diffusional properties from ab initio molecular dynamics simulations. *npj Comput. Mater.* **4**, 18 (2018).

5. Zhang, L., Han, J., Wang, H., Car, R. & E, W. Deep Potential Molecular Dynamics: A Scalable Model with the Accuracy of Quantum Mechanics. *Phys. Rev. Lett.* **120**, 143001 (2018).

6. Schmidt, J., Marques, M. R. G., Botti, S. & Marques, M. A. L. Recent advances and applications of machine learning in solid-state materials science. *npj Comput. Mater.* **5**, 83 (2019).

7. Friederich, P., Häse, F., Proppe, J. & Aspuru-Guzik, A. Machine-learned potentials for next-generation matter simulations. *Nat. Mater.* **20**, 750–761 (2021).

8. Behler, J. Four Generations of High-Dimensional Neural Network Potentials. *Chem. Rev.* **121**, 10037–10072 (2021).

9. Unke, O. T. *et al.* Machine Learning Force Fields. *Chem. Rev.* **121**, 10142–10186 (2021).

10. Deringer, V. L., Caro, M. A. & Csányi, G. Machine Learning Interatomic Potentials as Emerging Tools for Materials Science. *Adv. Mater.* **31**, 1–16 (2019).

11. Bartók, A. P., Payne, M. C., Kondor, R. & Csányi, G. Gaussian Approximation Potentials: The Accuracy of Quantum Mechanics, without the Electrons. *Phys. Rev. Lett.* **104**, 136403 (2010).

12. Bartók, A. P., Kondor, R. & Csányi, G. On representing chemical environments. *Phys. Rev. B* **87**, 184115 (2013).

13. Behler, J. & Parrinello, M. Generalized Neural-Network Representation of High-Dimensional Potential-Energy Surfaces. *Phys. Rev. Lett.* **98**, 146401 (2007).

14. Behler, J. Atom-centered symmetry functions for constructing high-dimensional neural network potentials. *J. Chem. Phys.* **134**, 074106 (2011).

15. Thompson, A. P., Swiler, L. P., Trott, C. R., Foiles, S. M. & Tucker, G. J. Spectral neighbor analysis method for automated generation of quantum-accurate interatomic potentials. *J. Comput. Phys.* **285**, 316–330 (2015).

16. Shapeev, A. V. Moment Tensor Potentials: A Class of Systematically Improvable Interatomic Potentials. *Multiscale Model. Simul.* **14**, 1153–1173 (2016).

17. Ko, T. W., Finkler, J. A., Goedecker, S. & Behler, J. A fourth-generation high-dimensional neural network potential with accurate electrostatics including non-local charge transfer. *Nat. Commun.* **12**, 398 (2021).

18. Zubatiuk, T. & Isayev, O. Development of Multimodal Machine Learning Potentials: Toward a Physics-Aware Artificial Intelligence. *Acc. Chem. Res.* **54**, 1575–1585 (2021).

19. Zong, H., Pilania, G., Ding, X., Ackland, G. J. & Lookman, T. Developing an

interatomic potential for martensitic phase transformations in zirconium by machine learning. *npj Comput. Mater.* **4**, 48 (2018).

20. Vandermause, J. *et al.* On-the-fly active learning of interpretable Bayesian force fields for atomistic rare events. *npj Comput. Mater.* **6**, 20 (2020).

21. Kailkhura, B., Gallagher, B., Kim, S., Hiszpanski, A. & Han, T. Y.-J. Reliable and explainable machine-learning methods for accelerated material discovery. *npj Comput. Mater.* **5**, 108 (2019).

22. Wen, M. & Tadmor, E. B. Uncertainty quantification in molecular simulations with dropout neural network potentials. *npj Comput. Mater.* **6**, 124 (2020).

23. Botu, V. & Ramprasad, R. Learning scheme to predict atomic forces and accelerate materials simulations. *Phys. Rev. B* **92**, 094306 (2015).

24. Fu, X. *et al.* Forces are not Enough: Benchmark and Critical Evaluation for Machine Learning Force Fields with Molecular Simulations. 1–25 (2022).

25. Cusentino, M. A., Wood, M. A. & Thompson, A. P. Explicit Multielement Extension of the Spectral Neighbor Analysis Potential for Chemically Complex Systems. *J. Phys. Chem. A* **124**, 5456–5464 (2020).

26. Bartók, A. P., Kermode, J., Bernstein, N. & Csányi, G. Machine Learning a General-Purpose Interatomic Potential for Silicon. *Phys. Rev. X* **8**, 041048 (2018).

27. Zhang, L. *et al.* End-to-end Symmetry Preserving Inter-atomic Potential Energy Model for Finite and Extended Systems. *Adv. Neural Inf. Process. Syst.* **32**, 4441–4451 (2018).

28. Babaei, H., Guo, R., Hashemi, A. & Lee, S. Machine-learning-based interatomic potential for phonon transport in perfect crystalline Si and crystalline Si with vacancies. *Phys. Rev. Mater.* **3**, 074603 (2019).

29. Qi, J. *et al.* Bridging the gap between simulated and experimental ionic conductivities in lithium superionic conductors. *Mater. Today Phys.* **21**, 100463 (2021).

30. Wu, E. A. *et al.* A stable cathode-solid electrolyte composite for high-voltage, long-cycle-life solid-state sodium-ion batteries. *Nat. Commun.* **12**, 1256 (2021).

31. Leung, W.-K., Needs, R. J., Rajagopal, G., Itoh, S. & Ihara, S. Calculations of Silicon Self-Interstitial Defects. *Phys. Rev. Lett.* **83**, 2351–2354 (1999).

32. Maroudas, D. & Brown, R. A. Atomistic calculation of the self-interstitial diffusivity in silicon. *Appl. Phys. Lett.* **62**, 172–174 (1993).

33. Tang, M., Colombo, L., Zhu, J. & Diaz de la Rubia, T. Intrinsic point defects in crystalline silicon: Tight-binding molecular dynamics studiesof self-diffusion, interstitial-vacancy recombination, and formation volumes. *Phys. Rev. B* **55**, 14279–14289 (1997).

34. Yang, M., Bonati, L., Polino, D. & Parrinello, M. Using metadynamics to build neural network potentials for reactive events: the case of urea decomposition in water. *Catal. Today* **18**, (2021).

35. Pant, S., Smith, Z., Wang, Y., Tajkhorshid, E. & Tiwary, P. Confronting pitfalls of AI-augmented molecular dynamics using statistical physics. *J. Chem. Phys.* **153**, 234118 (2020).

36. Desai, S., Reeve, S. T. & Belak, J. F. Implementing a neural network interatomic model with performance portability for emerging exascale architectures. *arXiv:2002.00054.* (2020).

37. Podryabinkin, E. V. & Shapeev, A. V. Active learning of linearly parametrized interatomic potentials. *Comput. Mater. Sci.* **140**, 171–180 (2017).

38. Thompson, A. P. *et al.* LAMMPS - a flexible simulation tool for particle-based materials modeling at the atomic, meso, and continuum scales. *Comput. Phys. Commun.* **271**, 108171 (2022).

39. Ward, L., Agrawal, A., Choudhary, A. & Wolverton, C. A general-purpose machine learning framework for predicting properties of inorganic materials. *npj Comput. Mater.* **2**, 16028 (2016).

40. Wang, H., Zhang, L., Han, J. & E, W. DeePMD-kit: A deep learning package for many-body potential energy representation and molecular dynamics. *Comput. Phys. Commun.* **228**, 178–184 (2018).

41. Kresse, G. & Furthmüller, J. Efficient iterative schemes for ab initio total-energy calculations using a plane-wave basis set. *Phys. Rev. B - Condens. Matter Mater. Phys.* **54**, 11169–11186 (1996).

42. Perdew, J. P., Ernzerhof, M. & Burke, K. Rationale for mixing exact exchange with density functional approximations. *J. Chem. Phys.* **105**, 9982–9985 (1996).

43. Ong, S. P. *et al.* Python Materials Genomics (pymatgen): A robust, open-source python library for materials analysis. *Comput. Mater. Sci.* **68**, 314–319 (2013).

44. Jain, A. *et al.* A high-throughput infrastructure for density functional theory calculations. *Comput. Mater. Sci.* **50**, 2295–2310 (2011).

45. He, X., Zhu, Y. & Mo, Y. Origin of fast ion diffusion in super-ionic conductors. *Nat. Commun.* **8**, 15893 (2017).

46. Wang, Y. *et al.* Design principles for solid-state lithium superionic conductors. *Nat.*

*Mater.* **14**, 1026–1031 (2015).

47. Virtanen, P. *et al.* SciPy 1.0: fundamental algorithms for scientific computing in Python. *Nat. Methods* **17**, 261–272 (2020).

**Figure captions**

**Fig. 1 | Testing of MLIPs.** The comparison of the atomic environments from (a) the interstitial-RE testing set $\mathcal{D}_{\text{RE-I}}^{\text{Testing}}$ (blue), (b) the vacancy-RE testing set $\mathcal{D}_{\text{RE-V}}^{\text{Testing}}$ (red), and (c) the original training dataset from Ref.[1] (cyan), using 6500, 6300, and 13553 atomic environments respectively. All atomic environments here are quantified by the 1st and the 2nd principal components from principal component analysis (PCA) of the SOAP descriptors. Comparison of the (g, h) atomic forces and (d, e) energies predicted from MLIP (GAP) versus the benchmark (DFT K4) on (e, h) the vacancy-RE testing set $\mathcal{D}_{\text{RE-V}}^{\text{Testing}}$ and (d, g) the interstitial-RE testing set $\mathcal{D}_{\text{RE-I}}^{\text{Testing}}$. Phonon dispersion of (f) the bulk Si and (i) the Si supercell with a single vacancy calculated by DFT K4 and GAP.

**Fig. 2 | Diffusions of point defects in Si.** Arrhenius plots of the diffusivity of (a) vacancy and (b) interstitial in Si from AIMD and MLIP-MD simulations. The missing data points indicate the failure of the MD simulations due to either the melting of the crystal structure or an insufficient number of atom hops to quantify diffusivities (See Supplementary Table 4, 5). The error bars of diffusivities are estimated based on the scheme in Ref. [4] (Methods).

**Fig. 3 | Si interstitials by MLIPs**. Atomistic configurations of (a) split-<110>, (b) tetrahedral, and (c) hexagonal Si interstitials. Comparison of DFT (red dashed lines) and MLIPs on (d) the formation energies $E_f$ of the Si interstitials and (e) the occurrence frequencies in AIMD (K2) and MLIP-MD simulations at 1230K (Methods).

**Fig. 4 | Errors in atom vibrations.** (a) Illustration of a vibrating Si atom by the distance $r_s$ to its static site and the distance $r_v$ to the nearest static vacancy site. (b) The potential energy surface (PES) of the vibrating Si atom moving along the direction (green arrow) towards the vacancy. The probability density (c) the nearest neighbor (NN) atoms and (d) non-NN atoms of the vacancy plotted as a function of $r_s$. (e-h) The probability density for the NN atoms of the vacancy plotted as a function of $r_v$ and $r_s$. All distributions are obtained from MD simulations at 1230K, using an averaging scheme described in the Methods.

**Fig. 5 | Errors of atomic forces.** (a) Illustration of force magnitude error $\delta_F$ and force direction $\delta_\theta$ of atomic force $F_{predicted}$ predicted by an MLIP in comparison to $F_{DFT}$ by DFT. For the RE atoms in the $\mathcal{D}_{RE-I}^{Testing}$, the errors (b) $\delta_F$ and (g) $\delta_\theta$ of DeePMD (orange) and GAP$_{PRX}$ (cyan), the corresponding distribution of (c) $\delta_F$ and (h) $\delta_\theta$, and the cumulative distribution function (CDF) of (d) $\delta_F$ and (i) $\delta_\theta$. The CDF of (e) $\delta_F$ and (j) $\delta_\theta$ on vacancy RE atoms in the $\mathcal{D}_{RE-V}^{Testing}$ set. Interstitial- and vacancy-RE atoms are atoms in the middle of migration identified in the $\mathcal{D}_{RE-I}^{Testing}$ and $\mathcal{D}_{RE-V}^{Testing}$ datasets, respectively.

**Fig. 6 | The performance of RE-enhanced MLIPs.** (a) Comparison of MLIPs for their force performance score $P(\mathcal{D}_{RE-I}^{Testing})$ on interstitial RE atoms versus the fitted activation energy $E_a^I$ of interstitial diffusion in MLIP-MD simulations. Black dashed lines are $E_a^I$ from AIMD simulations, and the red dashed line indicates $P(\mathcal{D}_{RE-I}^{Testing})$ calculated by DFT K1. (b) The calculated $E_a^I$ and $D_0^I$ of interstitial diffusion from AIMD simulations (blue bar and red dash line), original MLIPs (grey), and interstitial-enhanced MLIPs (green). (c) Comparison of MLIPs for their force accuracies $P(\mathcal{D}_{RE-I}^{Testing})$ on interstitial RE dataset versus $P(\mathcal{D}_{RE-V}^{Testing})$ on vacancy RE dataset, showing the Pareto

fronts (dash lines). These MLIPs are comprised of newly trained MLIPs by RE-enhanced training data with interstitials RE-I (cross) or vacancies RE-V (triangles), original MLIPs (circles), or MLIPs re-trained by the original dataset in Ref.[1] and selected by the RE-validation process (Methods). Thin black dashed lines are $P(\mathcal{D})$ calculated by DFT K1.

**Fig. 7 | Process of MLIP training and developing metrics.** (a) Conventional process of MLIP training including the validation process of fine-tuning hyperparameters. (b) The process of developing evaluation metrics.

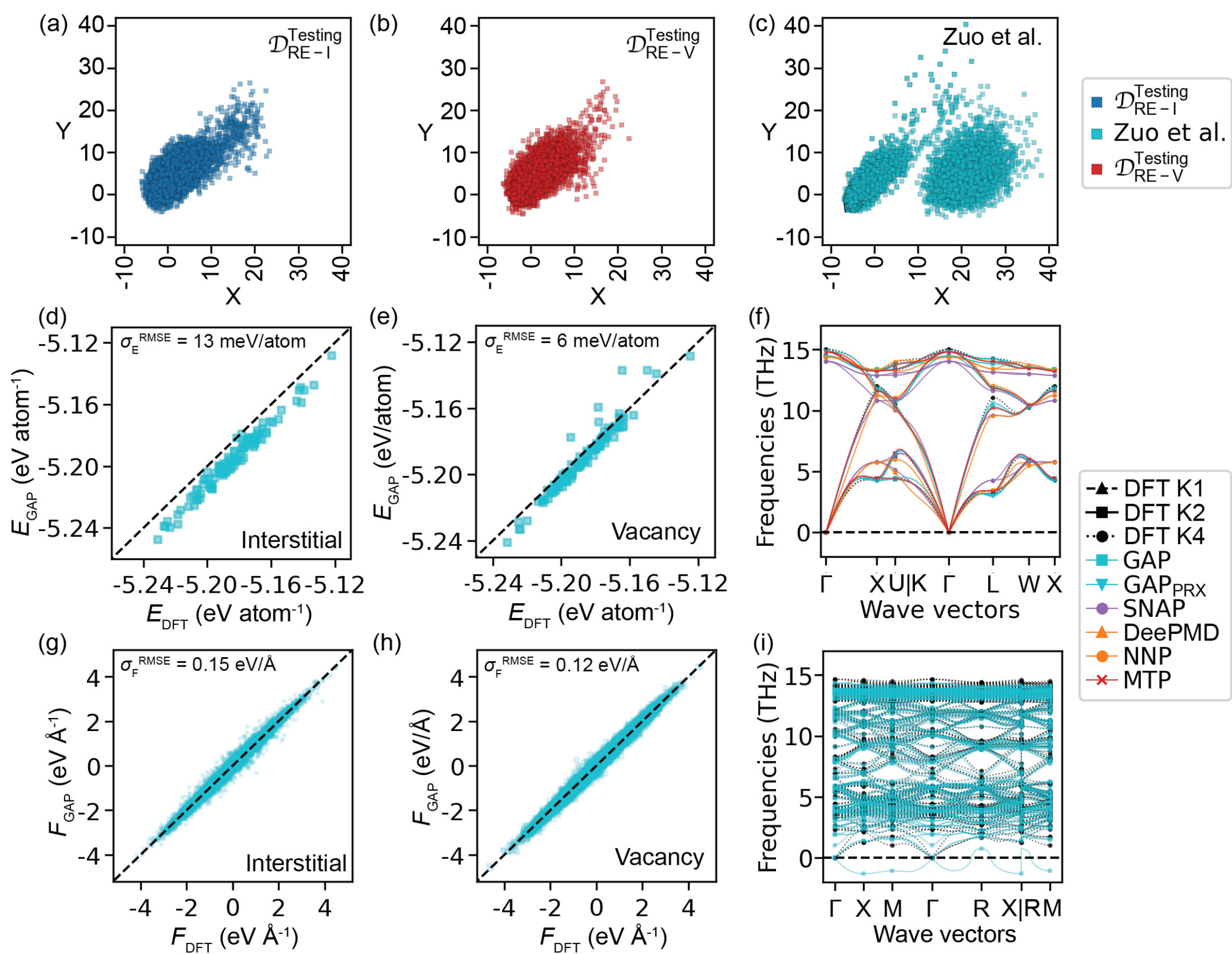

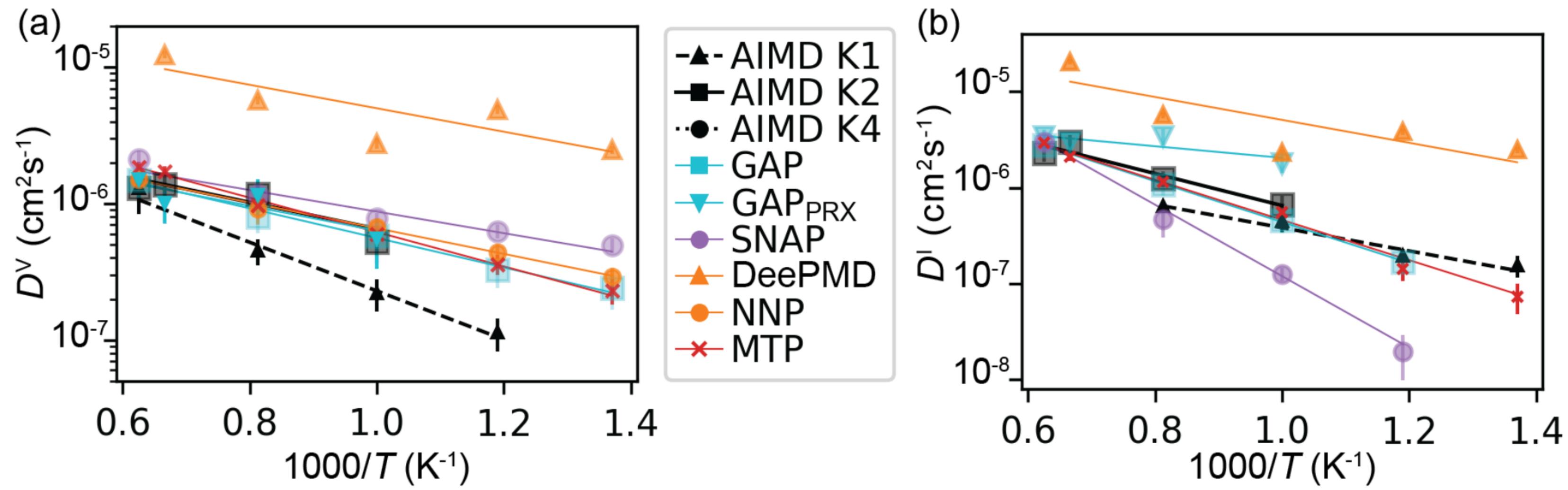

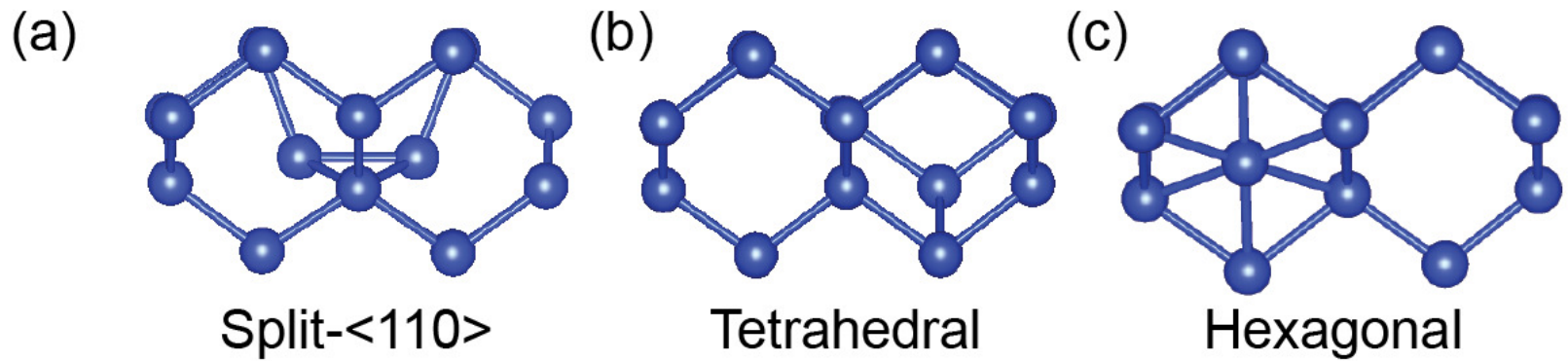

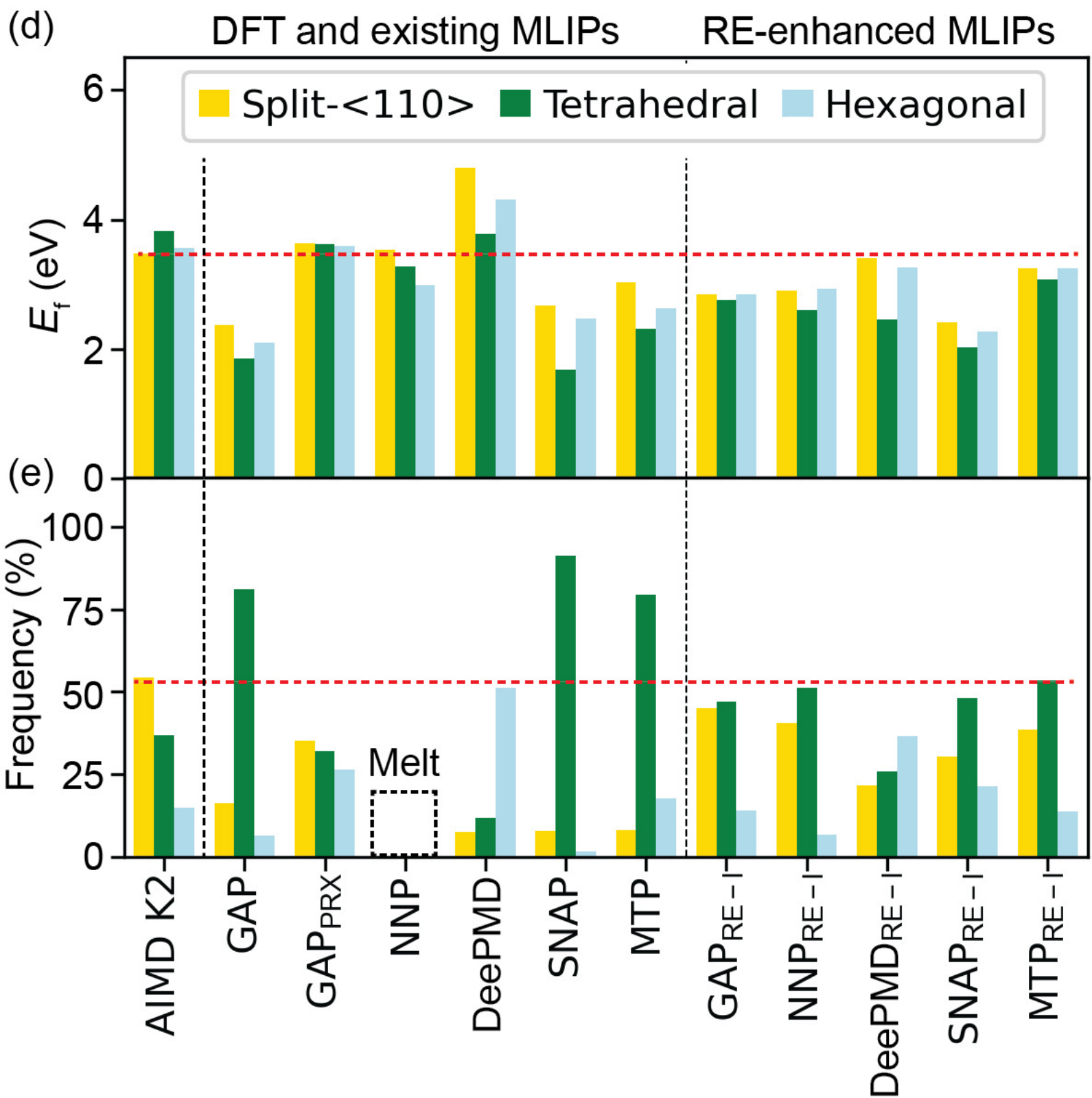

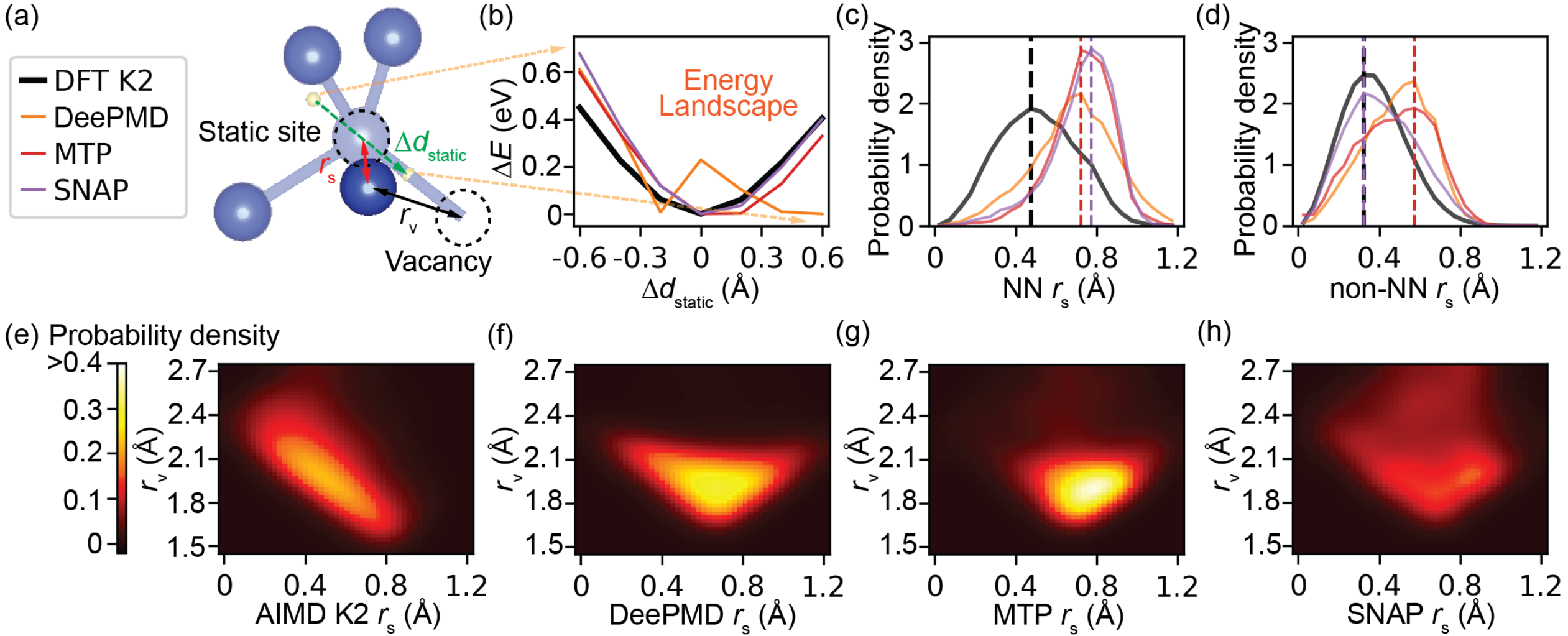

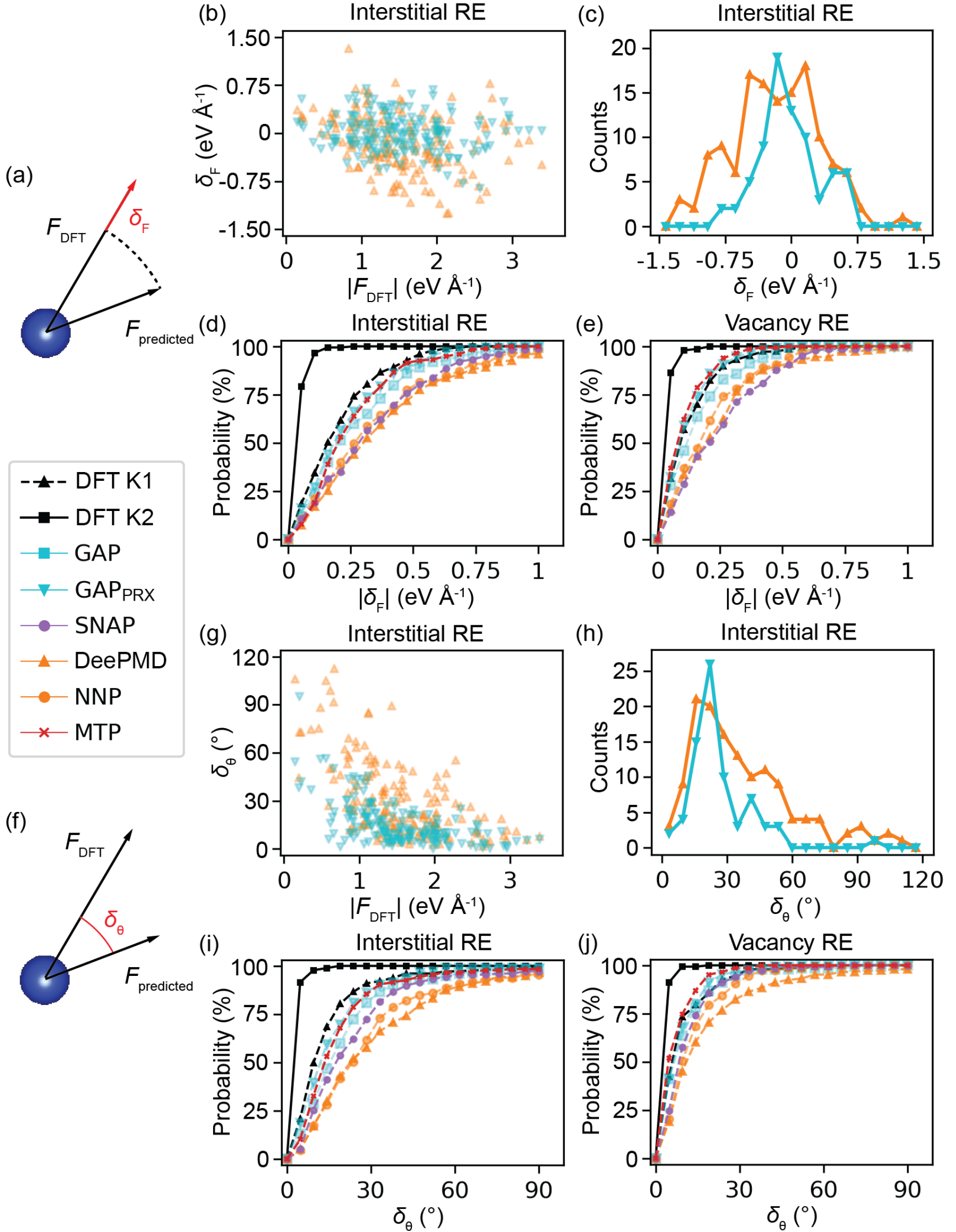

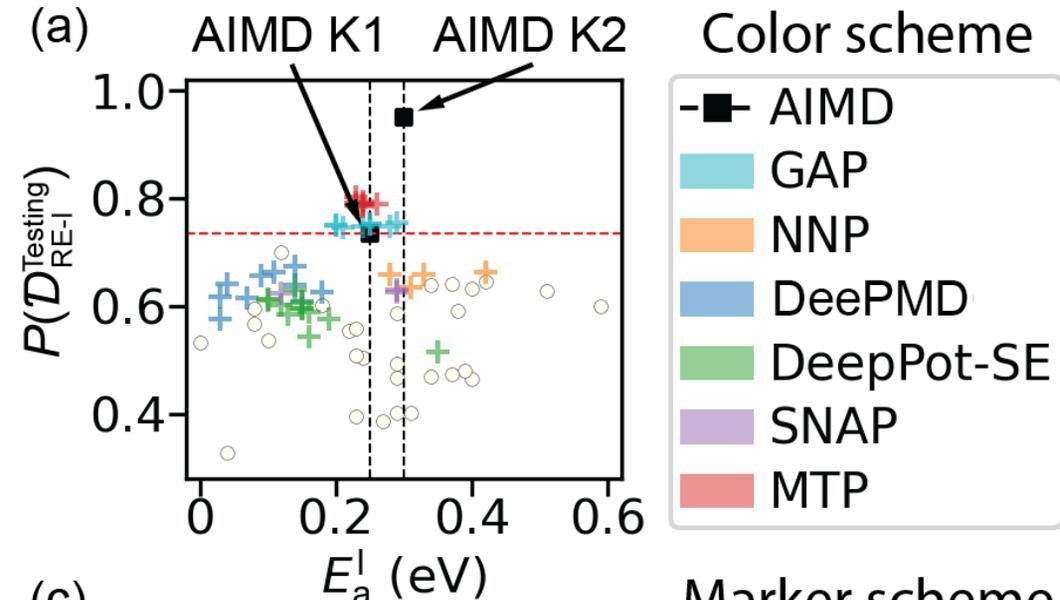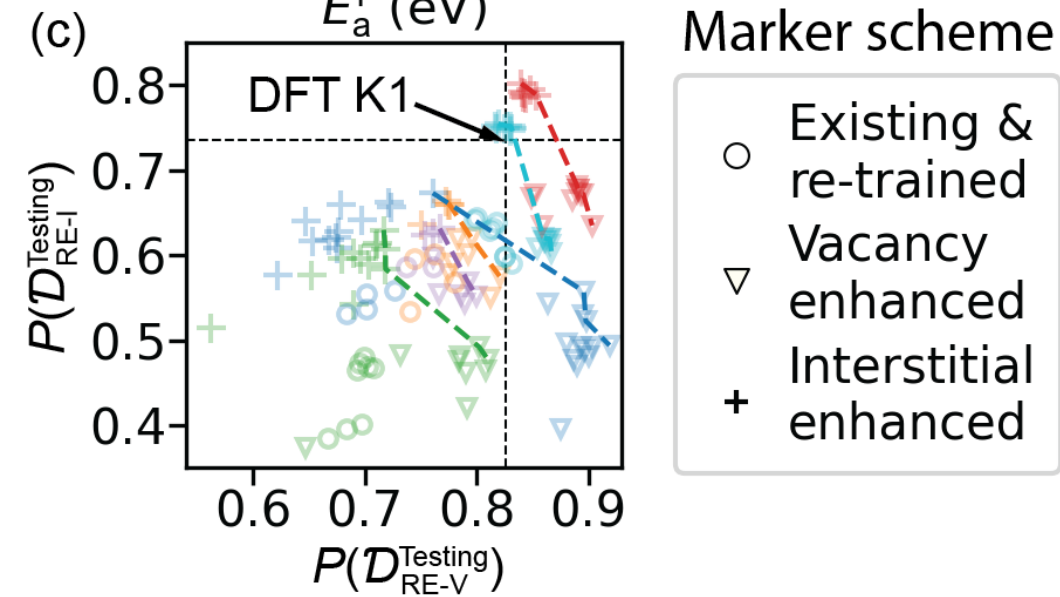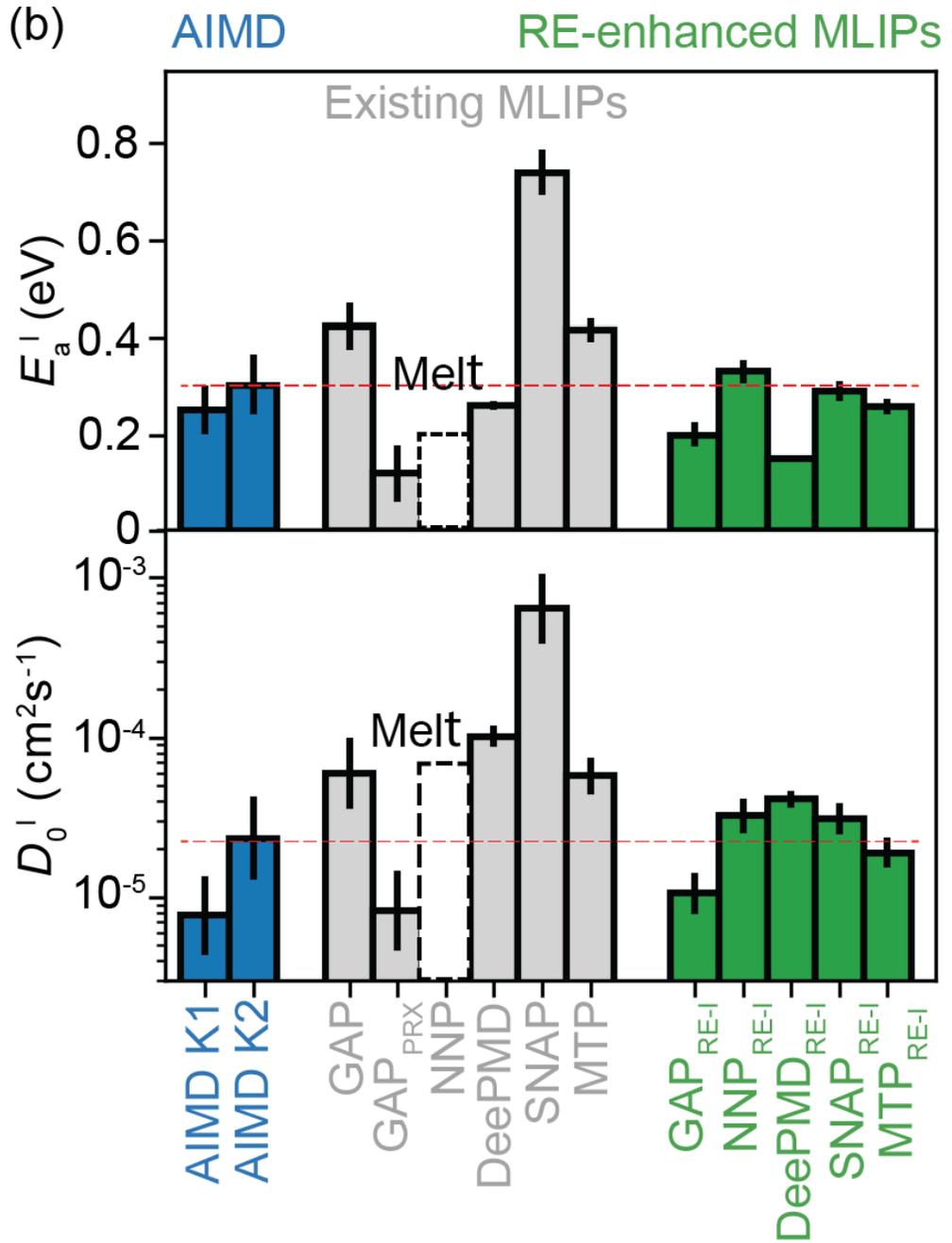

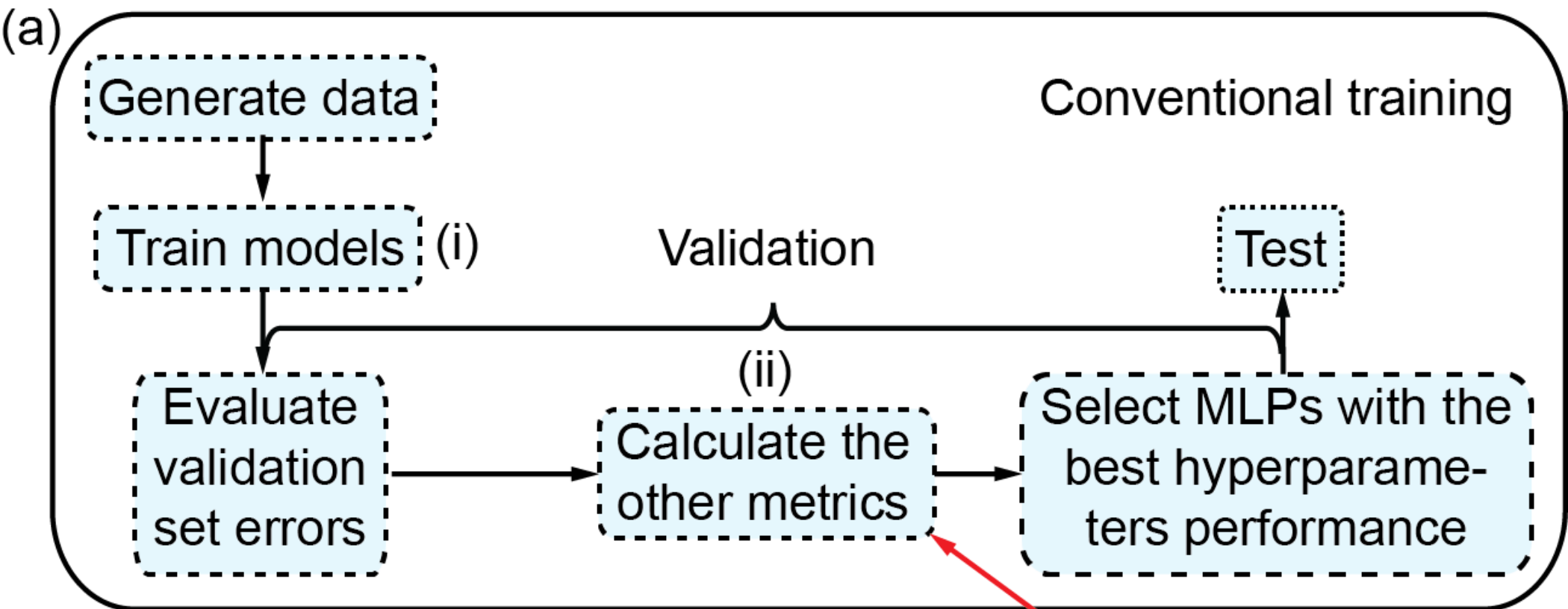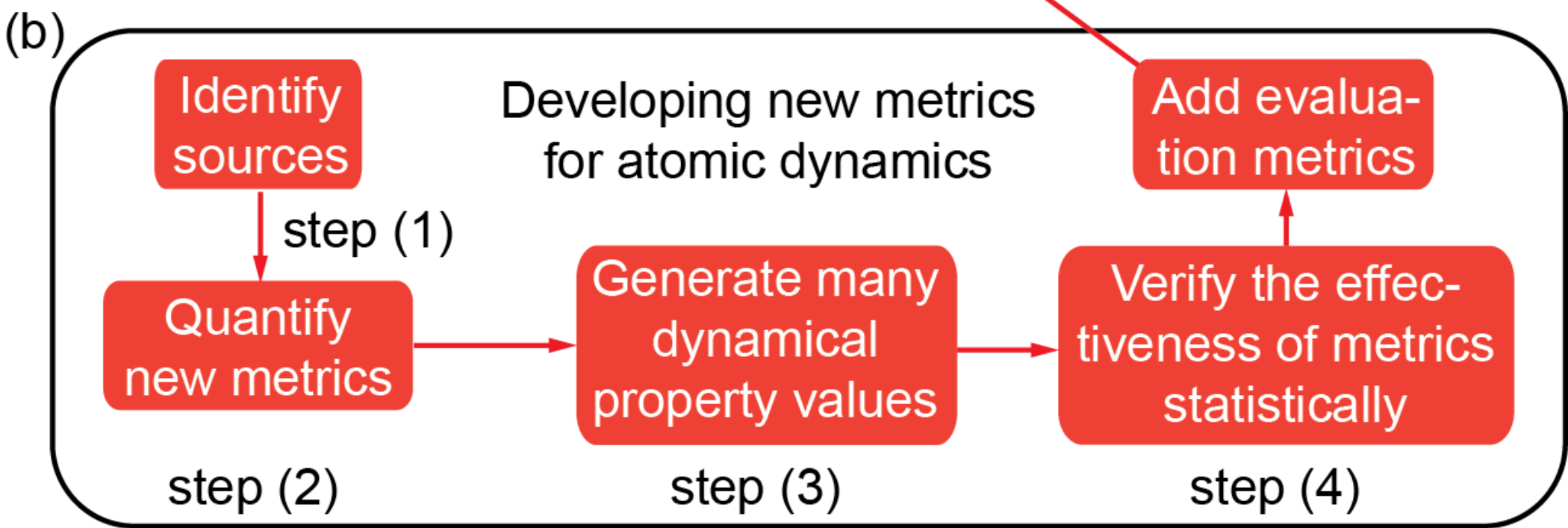